\documentstyle[twoside,fleqn,espcrc2igor]{article}

\input{psfig}

\newcommand{\lsim}{\mathrel{\mathop{\kern 0pt \rlap
  {\raise.2ex\hbox{$<$}}}
  \lower.9ex\hbox{\kern-.190em $\sim$}}}
\newcommand{\gsim}{\mathrel{\mathop{\kern 0pt \rlap
  {\raise.2ex\hbox{$>$}}}
  \lower.9ex\hbox{\kern-.190em $\sim$}}}
\newcommand{\gagamma}{g_{a\gamma\gamma}}

\newcommand{\AmS}{{\protect\the\textfont2
  A\kern-.1667em\lower.5ex\hbox{M}\kern-.125emS}}

\title{Prospects for solar axions searches with crystals via Bragg scattering}
\author{ I. G. Irastorza\thanks{Attending speaker:
         Igor.Irastorza@posta.unizar.es},
         S. Cebri\'{a}n, E. Garc\'{\i}a,
         D. Gonz\'{a}lez, A. Morales, J. Morales, A. Ortiz de
         Sol\'{o}rzano,\\ A. Peruzzi,
         J. Puimed\'{o}n, M. L. Sarsa,
         S. Scopel, J. A. Villar\\
         \vspace{4mm}
         Laboratorio de F\'{\i}sica Nuclear. Universidad de Zaragoza\\
         50009, Zaragoza, SPAIN
         }

\begin{document}

\begin{abstract}
A calculation of the expected signal due to Primakov coherent
conversion of solar axions into photons via Bragg scattering in
several solid--state detectors is presented and compared with
present and future experimental sensitivities. The axion window
$m_a\gsim 0.03$ eV (not accessible at present by other techniques)
could be explored in the foreseeable future with crystal detectors
to constrain the axion--photon coupling constant
$g_{a\gamma\gamma}$ below the latest bounds coming from
helioseismology. On the contrary a positive signal in the
sensitivity region of such devices would imply revisiting other
more stringent astrophysical limits derived for the same range of
the axion mass. The application of this technique to the COSME
germanium detector which is taking data at the Canfranc
Underground Laboratory leads to a 95\% C.L. limit
$g_{a\gamma\gamma}\leq 2.8\times 10^{-9}$ GeV$^{-1}$.
\end{abstract}


\maketitle

\section{Introduction}

Introduced twenty years ago as the Nambu--Goldstone boson of the
Peccei--Quinn symmetry to explain in an elegant way CP
conservation in QCD \cite{PQ}, the axion is remarkably also one of
the best candidates to provide at least a fraction of the Non
Barionic Dark Matter of the Universe.

Axion phenomenology is determined by its mass $m_a$ which in turn
is fixed by the scale $f_a$ of the Peccei--Quinn symmetry breaking
\cite{raffelt_review}. No hint is provided by theory about where
the $f_a$ scale should be. A combination of astrophysical,
cosmological and nuclear physics constraints restricts the allowed
range of viable axion masses into a relatively narrow
window\cite{kolbturner}: $10^{-6}  {\rm eV} \lsim  m_a \lsim
10^{-2} {\rm eV}$ and $3 \;{\rm eV} \lsim m_a \lsim  20\;  {\rm
eV}$.

The physical process used in axion search experiments is the
Primakov effect. It makes use of the coupling $g_{a\gamma \gamma}$
between the axion field and the electromagnetic tensor and allows
for the conversion of the axion into a photon. This coupling
appears automatically in every axion model, and like all the other
axion couplings, it is proportional to
$m_a$\cite{kolbturner,raffelt_review}, $g_{a\gamma \gamma} \simeq
0.19 \; C_{a\gamma \gamma} \; (m_a/{\rm eV}) \; 10^{-9} \; {\rm
GeV}^{-1}$, where the constant $C_{a \gamma \gamma}$ depends on
the axion model considered. Two popular models
\cite{raffelt_review} are the GUT--DFSZ axion ($C_{a\gamma
\gamma}=0.75 \pm 0.08$) and the KSVZ axion ($C_{a\gamma
\gamma}=-1.92 \pm 0.08$). However, the possibility to build viable
axion models with different values of $C_{a\gamma \gamma}$ and the
theoretical uncertainties involved \cite{moroi} imply that a very
small or even vanishing $\gagamma$ cannot be in principle
excluded.

\section{Primakov conversion in crystals}

Axions can be efficiently produced in the interior of the Sun by
Primakov conversion of the blackbody photons in the fluctuating
electric field of the plasma. Solid state detectors provide a
simple mechanism for detecting these axions. Axions can pass in
the proximity of the atomic nuclei of the crystal where the
intense electric field can trigger their conversion into photons.
Due to the fact that the solar axion flux has an outgoing average
energy of about 4 keV (corresponding to the temperature in the
core of the Sun, $T\sim 10^7 K$) they can produce detectable
x--rays in a crystal detector. Depending on the direction of the
incoming axion flux with respect to the planes of the crystal
lattice, a coherent effect can be produced when the Bragg
condition is fulfilled, leading so to a strong enhancement of the
signal.

\begin{figure}[tb]
\mbox{\psfig{figure=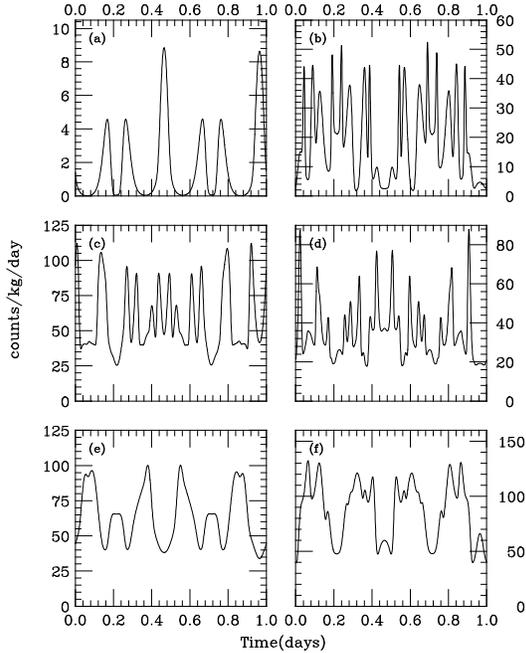,width=70mm}}
\caption{\small Expected axion signals for Primakov conversion in
various crystals as a function of time for $\gagamma=10^{-8}\;
{\rm
GeV}^{-1}$. 
From top--left to bottom--right: a) Ge, 2 keV$\leq E_{ee}\leq$2.5
keV; b) Ge, 4 keV$\leq E_{ee}\leq$4.5 keV; c) TeO$_2$, 5 keV$\leq
E_{ee}\leq$7 keV; d) TeO$_2$, 7 keV$\leq E_{ee}\leq $9 keV; e)
NaI, 2 keV$\leq E_{ee}\leq $4 keV; f) NaI, 4 keV$\leq E_{ee}\leq$6
keV.} \label{fig:axions}
\end{figure}

Making use of the calculation of the flux of solar axions of Ref.
\cite{raffelt}, as well as the cross--section of the process and
appropriate cristallographic information, we calculate the
expected axion-to-photon conversion count rate in a solid-state
detector (See ref. \cite{axiones_canfranc} for further details).
Some examples of this count rate for several materials and energy
windows are shown in figure \ref{fig:axions} as a function of time
during one day. As expected, the signal presents a strong
sub--diary dependence on time, due to the motion of the Sun in the
sky. All this information can be used\cite{axiones_canfranc} to
extract a posible axion signal from a set of experimental data or,
in case of a non--appearance of such a signal, to obtain a limit
on the axion-photon coupling $\gagamma$. This process is
independent of $m_a$ and so are the achievable bounds for
$\gagamma$. This fact is particularly appealing, since other
experimental techniques are limited to a more restricted mass
range \cite{cavities,tokio}.

The method described above has been applied to the 311 days of
data obtained by the COSME 0.234 kg germanium detector (which is
also being used for Dark Matter detection, as is briefly commented
in the Dark Matter review talk in these proceedings) in the
Canfranc Underground Laboratory, with a effective threshold of 2.5
keV and a low energy background of 0.7 c/keV/kg/day. With these
conditions and despite its lower statistics, we reach a limit
$\gagamma\lsim 2.8 \times 10^{-9}\; {\rm GeV}^{-1}$ very close to
the one obtained by the SOLAX Collaboration \cite{cosmesur} which
is the (mass independent but solar model dependent) most stringent
laboratory bound for the axion--photon coupling obtained so far.

\section{Future prospects}

\begin{table*}[hbt]

\setlength{\tabcolsep}{0.5pc}
\newlength{\digitwidth} \settowidth{\digitwidth}{\rm 0}
\catcode`?=\active \def?{\kern\digitwidth}
\caption{\small Axion search sensitivities for running
(COSME,DAMA), being installed (CUORICINO, ANAIS) and planned
(CUORE, GENIUS) experiments are compared to the result of
SOLAX\protect\cite{cosmesur} (See \cite{axiones_canfranc} for
references). A Pb detector is also included (see text). The
coefficient $K$ is defined in Eq.(\ref{eq:limit}).}

\label{tab:exp}
\small
\begin{tabular*}{\textwidth}{@{}l@{\extracolsep{\fill}}rrrrrr}
\hline

                 & \multicolumn{1}{r}{{\bf K}}
                 & \multicolumn{1}{r}{{\bf M} (kg)}
                 & \multicolumn{1}{r}{{\bf b} (cpd/kg/keV)}
                 & \multicolumn{1}{r}{{\bf E$_{th}$} (keV)}
                 & \multicolumn{1}{r}{{\bf FWHM} (keV)}
                 & \multicolumn{1}{r}{ $\gagamma^{lim}(2\; {\rm years})$ (GeV$^{-1}$)}       \\
\hline
\small
{\bf Ge}\cite{cosmesur} & 2.5 & 1 & 3 & 4 & 1 & 2.7$\times
10^{-9}$\\ {\bf Ge} & 2.3 & 0.234 & 0.7 & 3 & 0.4 & 2.4$\times
10^{-9}$\\ {\bf Ge} & 2.5 & 1000 & 1$\times 10^{-4}$ & 4 & 1 &
3$\times 10^{-10}$\\ {\bf TeO$_2$} & 3 & 42 & 0.1 & 5 & 2 &
1.3$\times 10^{-9}$ \\ {\bf TeO$_2$}& 2.8 & 765 & $1\times
10^{-2}$ & 3 & 2 & 6.3$\times 10^{-10}$ \\ {\bf NaI} & 2.7 & 87 &
1 & 2 & 2 & 1.4$\times 10^{-9}$ \\ {\bf NaI} & 2.8 & 107 & 2 & 4 &
2 & 1.6$\times 10^{-9}$ \\ {\bf Pb} & 2.1 & 1000 & $1\times
10^{-4}$ & 4 & 1 & 2.5$\times 10^{-10}$ \\ \hline

\end{tabular*}
\end{table*}





The sensitivity of an axion experimental search can be expressed
as the upper bound of $\gagamma$ which such experiment would
provide from the non--appearance of the axion signal, for a given
crystal, background and exposure. It is easy to verify that the
ensuing limit on the axion--photon coupling $\gagamma^{lim}$
scales with the background and exposure in the following way:

\begin{eqnarray}
\gagamma\leq \gagamma^{lim}\simeq K \left(\frac{\rm b}{{\rm
cpd/kg/keV}}\times\frac{\rm kg} {{\rm M}}\times\frac{\rm
years}{\rm T} \right)^{\frac{1}{8}} \nonumber \\
 \times 10^{-9} \; {\rm GeV}^{-1}
\label{eq:limit}
\end{eqnarray}
\noindent where $M$ is the total mass and $b$ is the average
background. The factor $K$ depends on the parameters of the
crystal, as well as on the experimental threshold and resolution.

 In order to perform a systematic analysis of the
axion--detection capability of crystal detectors, we have applied
the technique described in the previous section to several
materials. The result is summarized in Table \ref{tab:exp}, where
the limit given by the experiment of Ref.\cite{cosmesur} is
compared to those attainable with COSME and other running, being
installed and planned crystal detector experiments (See
\cite{axiones_canfranc} for references).

In Table \ref{tab:exp} a Pb detector is also included, to give an
indication of the best improvement that one would expect by
selecting heavy materials to take advantage of the proportionality
to $Z^2$ of the cross section (crystals of PbWO$_4$ have been
considered in the setup of CUORE).


\begin{figure}[tb]
\psfig{figure=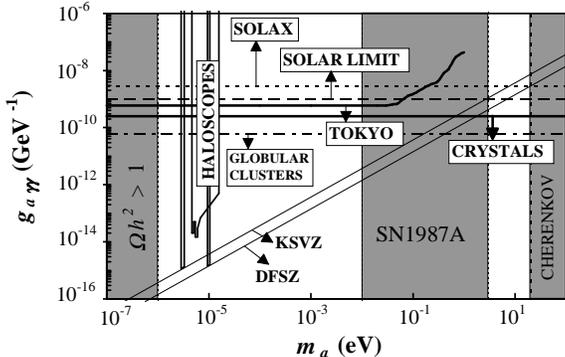,width=70mm}
\caption{\small The solar axion limit attainable with crystal
detectors (horizontal thick line) is compared to the present
astrophysical and experimental bounds and to the DFSZ and KSVZ
axion theoretical predictions.}
\label{fig:axionlimits}

\end{figure}


In Fig.\ref{fig:axionlimits} the result of our analysis is
compared to the present astrophysical and experimental bounds in
the plane $m_a$--$\gagamma$. The horizontal thick line represents
the constraint $\gagamma\lsim$3$\times$10$^{-10}$ GeV$^{-1}$ taken
from Table \ref{tab:exp}. The mass intervals $m_a\lsim 10^{-6}$
eV, $10^{-2}$ eV $\lsim m_a \lsim 3$ eV, $m_a\gsim 20$ eV are
excluded respectively by cosmological
overclosure\cite{kolbturner}, SN1987A and oxygen excitation in
water Cherenkov detectors, the edges of all the excluded regions
being subject to many astrophysical and theoretical uncertainties
(See \cite{axiones_canfranc} for further details about the
figure).

As shown in the expression of the $\gagamma$ bound of
Eq.(\ref{eq:limit}) the improvement in background and accumulation
of statistics is washed out by the 1/8 power dependence of
$\gagamma$ on such parameters. It is evident, then, that crystals
have no realistic chances to challenge the globular cluster limit.
A discovery of the axion by this technique would presumably imply
either a systematic error in the stellar--count observations in
globular clusters or a substantial change in the theoretical
models that describe the late--stage evolution of low--metallicity
stars.
 On the other hand, the sensitivity required for
crystal--detectors in order to explore a range of $\gagamma$
compatible with the solar limit \cite{axiones_canfranc}, appears
to be within reach, provided that large improvements of background
as well as substancial increase of statistics be guaranteed.


\begin{thebibliography}{10}

\small
\bibitem{PQ}
R. D. Peccei and H. R. Quinn, {\em Phys. Rev. Lett.} {\bf 38}
(1977) 1440.

\bibitem{kolbturner}
E. W. Kolb and M. S. Turner, {\em The Early Universe}, Addison
Wesley
  Publishing, (1990).

\bibitem{cosmesur}
The SOLAX Collaboration, {\em Nucl. Phys.} {\bf B}(Proc.
Suppl.){\bf 70} (1999) 59.

\bibitem{cavities}
C. Hagmann et al., {\em Phys. Rev.
  Lett.} {\bf 80} (1998) 2043.

\bibitem{tokio}
S. Moriyama et al., {\em
  Phys. Lett.} {\bf B434} (1998) 147.

\bibitem{raffelt_review}
G. Raffelt, {\em Phys. Rep.} {\bf 198} (1990) 1.




\bibitem{moroi}
T. Moroi and H. Murayama, {\em Phys. Lett.} {\bf B440} (1998)
69-76.

\bibitem{raffelt}
K. van Bibber, P.M. McIntyre, D.E. Morris and G. Raffelt, {\em
Phys. Rev.} {\bf
  D39} (1989) 2089.

\bibitem{axiones_canfranc}
S. Cebri\'{a}n, E. Garc\'\i a, D. Gonz\'{a}lez, I. G. Irastorza, A.
Morales, J. Morales, A. Ortiz de Sol\'{o}rzano, J. Puimed\' on, A.
Salinas, M. L. Sarsa, S. Scopel, J. A. Villar, {\em Astrop. Phys.}
{\bf 10} (1999) 397-404, and references therein.


\end{thebibliography}
\end{document}